\documentclass[manuscript]{aastex}

\usepackage{natbib}
\usepackage{color}
 \usepackage{ulem}
\def\gsim{\mbox{{\scriptsize \raisebox{-.9ex}
      {$\;\stackrel{{\textstyle >}}{\sim}\,$} }} }

\slugcomment{Prepared for submission to Astrophys. J.}

\shorttitle{Neutrino Emissivities via Deuterons}
\shortauthors{Nasu et al.}

\begin{document}

\title{Neutrino Emissivities from Deuteron-Breakup and Formation in Supernovae}


\author{S. Nasu}
\affil{Department of Physics, Osaka University, Toyonaka, Osaka 560-0043, Japan}
\email{nasu@phys.sci.osaka-u.ac.jp}

\author{S. X. Nakamura}
\affil{Department of Physics, Osaka University, Toyonaka, Osaka 560-0043, Japan}
\email{nakamura@kern.phys.sci.osaka-u.ac.jp}

\author{K. Sumiyoshi}
\affil{Numazu College of Technology, Ooka 3600, Numazu, Shizuoka 410-8501, Japan}
\email{sumi@numazu-ct.ac.jp}

\author{T. Sato}
\affil{Department of Physics, Osaka University, Toyonaka, Osaka 560-0043, Japan}
\email{tsato@phys.sci.osaka-u.ac.jp}

\author{F. Myhrer}
\affil{Department of Physics and Astronomy, University of South Carolina, Columbia SC 29208, USA}
\email{myhrer@physics.sc.edu}

\and

\author{K. Kubodera}
\affil{Department of Physics and Astronomy, University of South Carolina, Columbia SC 29208, USA}
\email{kubodera@caprine.physics.sc.edu}




\begin{abstract}

Neutrino emissions from
electron/positron capture on the deuteron
and the nucleon-nucleon fusion processes
in the surface region of a supernova core are studied.
These weak processes are evaluated in the
standard nuclear physics approach, which consists of 
one-nucleon and two-nucleon-exchange currents and nuclear wave functions 
generated by a high precision nucleon-nucleon potential.
In addition to the cross sections 
for these processes involving the deuteron,
we present neutrino emissivities due to these processes calculated
for typical profiles of core-collapsed supernovae.
These novel neutrino emissivities are compared 
with the standard neutrino emission mechanisms.
We find that the neutrino emissivity due to the
electron capture on the deuteron 
is comparable to that on the proton in the deuteron abundant region.
The implications of the new channels involving deuterons
for the supernova mechanism are discussed.

\end{abstract}


\section{Introduction} \label{sec:introduction}

The neutrinos play pivotal roles in core-collapse supernovae 
and their subsequent evolution to neutron stars.  
Neutrino reactions in the dense matter 
of a supernova core 
are crucial for understanding the supernova explosion mechanism, 
which is still elusive despite extensive studies over decades.  
It is therefore essential to identify all neutrino processes, 
both neutrino-emission and neutrino-absorption processes, 
that can be important in the supernova environment. 
Failing to include all the relevant neutrino processes
in supernova modeling
may have significant consequences
for the theoretical understanding of the supernova explosion. 
The emission of neutrinos 
acts as a  cooling mechanism of the central core, 
and the addition of any neutrino emission channels
that have not been considered so far could
affect this cooling mechanism. 
A portion of the emitted neutrinos are subsequently absorbed 
by the material behind the shock wave and thereby act as a heating agent.  
Additional sources of neutrinos enhance this 
neutrino-heating mechanism and may help the revival of 
the  stalled shock wave and 
lead to a successful modeling of
 supernova explosion \citep{bet90,kot06,jan07}.  
The neutrinos emitted gradually ($\sim$20 s) 
from a nascent neutron star 
(proto-neutron star) in a supernova explosion,  
can be detected as supernova neutrinos at terrestrial 
neutrino detectors, like in the case of SN1987A \citep{suz94}, 
and 
can be a useful source of information
about the neutrino emission mechanisms.

Recent calculations have shown  
that deuterons, tritons and $^{3}$He 
can appear copiously in the regions 
between the supernova core and the shockwave 
\citep{sumroe08,arc08,hemp12}. 
These light elements have so far not been included 
in the tables of equation of state (EOS)~\citep{lat91,she98a,she98b} 
that are routinely used in supernova simulations 
where  
the nuclear species are limited to the proton, neutron, $^{4}$He 
and one ``representative heavy nucleus" that is assumed to
simulate the roles of all heavy nuclei.
The light elements with mass number $A\!=\!2\,{\rm and}\,3$ 
can be abundant in 
hot and moderately dense matter 
($  < 10^{13}g/cm^{3}$)
under nuclear statistical equilibrium~\citep{sumroe08,arc08,hemp12,furusawa13a} 
and should be considered in studying the supernova mechanism. 
They appear in the heating region behind a shockwave, 
and also in the cooling region at the surface of a proto-neutron star;
their appearance gives a new contribution
to the neutrino opacity.   
For example, ~\citet{sumroe08} showed 
in a snapshot 150 msec after the bounce 
that  the deuteron mass fraction 
amounts to about 10\%   
in the neutrino-emitting 
cooling regions at densities $\sim$10$^{11}$-10$^{12}$ g/cm$^{3}$. 
Neutrino processes in this cooling region 
are essential for determining the flux and spectra of emitted neutrinos, 
which in turn affect the efficiency of neutrino heating 
behind the shockwave. 
Nakamura et al.~\citep{nakamura09} 
investigated neutrino absorption on deuterons 
as an additional heating mechanism on top of the neutrino
reactions on nucleons and $^{4}$He, while
Arcones et al.~\citep{arc08} studied neutrino reactions on tritons and $^{3}$He 
to evaluate their influences on the neutrino spectra 
at the outer layer of a proto-neutron star.

According to Refs.~\citet{sumroe08,arc08},
the deuteron fraction can be
larger than the proton fraction 
in part of the neutrino-sphere region between the 
shock wave and the proto-neutron star surface.
This indicates that weak-interaction deuteron breakup may play a significant role  
in neutrino emission processes,  
possibly altering the conventional understanding of
the role of the protons in the neutrino-emission processes
as well as the neutronization of matter.  
An additional reaction to be investigated in this work is 
deuteron formation in nucleon-nucleon scattering,
which also leads to neutrino emission.
Although both these neutrino emission processes certainly exist 
on top of the conventional neutrino-emission processes,  
they have so far not been considered in supernova simulations. 

In this article we study neutrino emissions 
from \underline{d}euteron
\underline{b}reakup/\underline{f}ormation processes
(DBF for short)
in the surface region of a proto-neutron star,
where neutrino emissions act 
as a cooling mechanism.  
We present the first evaluation of the neutrino emissivities from
electron/position-capture on the deuteron (deuteron breakup)
and from the nucleon-nucleon weak fusion processes
(deuteron formation);
see (\ref{eqn:eled})-(\ref{eqn:pnd}) below.
The neutrino emissivities arising from DBF
will be compared with those coming from the ``conventional" processes;
by ``conventional" processes we mean 
the neutrino emission processes
which have been previously considered in the literature,
and which are listed in (\ref{eqn:elep})-(\ref{eqn:epem}) below. 
The neutrino emissivities due to DBF reported here 
are expected to be useful
for numerical simulations of supernova explosion 
and proto-neutron star cooling.  

Theoretical treatments of electroweak processes 
in two-nucleon systems 
are well developed.
For low-energy neutrino-deuteron reactions, serious 
efforts to reduce theoretical uncertainties
have been made in order to analyze data from the Sudbury Neutrino 
Observatory~\citep{nsk01,net02}.
One approach 
is the standard nuclear physics
approach (SNPA) that involves nuclear wave functions
derived from high-precision 
phenomenological nuclear potentials,
and one-nucleon and two-nucleon electroweak currents.
SNPA has been well tested by analyses of photo-reactions, electron scattering,
and muon capture on the two-nucleon systems~\citep{doi90,tamu92,sato95}.
Another theoretical approach,
effective field theory (EFT) 
consisting of
nucleons and pions, 
has been developed and applied to low-energy electroweak processes~\citep{kubo04}.
Both methods essentially agree with each other for
low-energy electroweak processes in the two-nucleon systems.
The $pp$-fusion process, $pp\to d e^- \nu_e$, 
is one of such processes relevant to this work. 
This reaction has been studied with both SNPA and EFT, 
and good agreement between the two methods
has been found~\citep{sch98,tsp03}.
Another nucleon-nucleon fusion process relevant to this work is 
neutron-neutron fusion, which was previously studied with EFT~\citep{ando06}.
In the present work we  
adopt SNPA.

This article is arranged as follows.  
In section \ref{sec:reaction}
we discuss neutrino emissions via DBF
relevant to the supernova environment.
The theoretical framework for calculating 
the cross sections for neutrino emission via DBF
and the corresponding neutrino emissivities are outlined in 
section \ref{sec:formulation}, 
and the numerical results are presented 
in sections \ref{sec:cros} and \ref{sec:SecEmissivity}.
The implications and a summary of these results
for the supernova mechanism are discussed 
in section~\ref{sec:summary}.    

\section{Neutrino Reactions Involving the Deuteron}\label{sec:reaction}

We consider neutrino emissions via deuteron breakup/formation (DBF):
\begin{eqnarray}
d   + e^- &\rightarrow&  n + n   +      \nu_e   \ , \label{eqn:eled} \\
d   + e^+ &\rightarrow&  p + p   + \bar{\nu}_e  \ , \label{eqn:posd} \\
n   + n   &\rightarrow&  d + e^- + \bar{\nu}_e  \ , \label{eqn:nnd}  \\
p   + p   &\rightarrow&  d + e^+ +      \nu_e   \ , \label{eqn:ppd}  \\
p   + n   &\rightarrow&  d + \nu + \bar{\nu}    \ . \label{eqn:pnd}
\end{eqnarray}
The reactions (\ref{eqn:eled}) and (\ref{eqn:posd}) are 
deuteron breakup via $e^-$/$e^+$-capture,
whereas   
the reactions (\ref{eqn:nnd}), (\ref{eqn:ppd}) and (\ref{eqn:pnd}) are 
deuteron formation through nucleon-nucleon scattering.
The first four reactions that are caused by the charged-current (CC), 
can only emit  $\nu_e$ or $\bar{\nu}_e$,
whereas the last reaction occurring via the neutral-current (NC)
gives rise to $\nu\bar{\nu}$ pair-emission 
of all three flavors.

The reactions (\ref{eqn:eled})-(\ref{eqn:pnd})  
are to be compared with the conventional
neutrino-emission reactions
that have been routinely included
in the study of supernovae and neutron stars. 
They are:
\begin{eqnarray}
p   + e^- &\rightarrow&  n   +      \nu_e   \ , \label{eqn:elep} \\
n   + e^+ &\rightarrow&  p   + \bar{\nu}_e  \ , \label{eqn:posn} \\
n   + n   &\rightarrow&  p + n + e^- + \bar{\nu}_e  \ , \label{eqn:nnnp}  \\
p   + p   &\rightarrow&  p + n + e^+ +      \nu_e   \ , \label{eqn:ppnp}  \\
N   + N'  &\rightarrow&  N + N' + \nu + \bar{\nu}  \ , \label{eqn:nnbrems}\\
e^- + e^+ &\rightarrow& \nu + \bar{\nu} \ . \label{eqn:epem}
\end{eqnarray}
Reactions (\ref{eqn:elep}) and (\ref{eqn:posn}) represent
the direct Urca processes, in which  
$e^-$/$e^+$-captures on a single nucleon
produce $\nu_e$/$\bar{\nu}_e$.
Reactions (\ref{eqn:nnnp}) and (\ref{eqn:ppnp}) are 
the modified Urca processes,
where nucleon-nucleon collisions  
lead to $\bar{\nu}_e$/$\nu_e$ emission.  
In the last two reactions a pair of $\nu\bar{\nu}$ of all flavors is produced; 
it is a common practice to refer to the process~(\ref{eqn:nnbrems}) 
as nucleon-nucleon bremsstrahlung  and 
the process~(\ref{eqn:epem}) as $e^+e^-$ annihilation.

The NC reactions produce pairs of $\nu\bar{\nu}$
and act as a cooling mechanism.
When the CC reactions take place frequently enough, 
the proton and neutron fractions are determined 
through $\beta$-equilibrium, $\mu_e 
= \mu_n -\mu_p\,  + \,\mu_\nu$.
Inside the proto-neutron star,
chemical equilibrium 
is realized among electrons, positrons, nucleons
and neutrinos.
At the surface of the proto-neutron star(densities
 $\sim 10^{11} -10^{13}$g/cm$^3$)
where neutrinos are not trapped,
one must solve the neutrino transfer equation 
with detailed information about the neutrino reaction rates,
in order to determine 
the neutrino distribution and its evolution 
 associated with the change of matter composition.
We note that the reaction rates depend 
on the degeneracy of leptons and nucleons
in the supernova environment;
the high degeneracy of the leptons  and/or nucleons
can significantly suppress the reaction rates.  
This makes it important to take a proper account of
the Pauli blocking factors for particles participating in the reactions.

\subsection{Deuteron Breakup via  Electron/Positron Capture} 
\label{sec:dbreakup}

The $e^-/e^+$-capture reactions on the deuteron, 
(\ref{eqn:eled}) and (\ref{eqn:posd}),
may  influence significantly 
the $\nu_e/\bar{\nu}_e$ emissivities
{\it at the  surface} of a proto-neutron star,
where deuterons could be abundant~\citep{sumroe08,arc08}.
In a supernova simulation that includes the deuteron abundance
(or, more generally, the light element abundance), 
there are fewer free nucleons and more bound nucleons
than in conventional supernova simulations.
Therefore, depending on whether 
the $e^-/e^+$-capture rates on the deuteron
are larger or smaller than those on the proton,
the net capture rate per proton (both free and bound protons counted)
is enhanced or suppressed 
by the deuteron abundance. 
As will be presented, we find 
the latter to be the case.   

Just like the first  direct Urca process (\ref{eqn:elep}), 
$e^-$-capture on the deuteron acts 
as a source of neutronization of the proto-neutron star and   
drives the dense matter toward the neutron-rich side 
by changing protons into neutrons with neutrino emissions.  
Similar to the second   direct Urca process (\ref{eqn:posn}), 
$e^+$-capture on the deuteron acts
as a counter reaction, changing neutrons into protons.  
In addition, 
in matter
with trapped neutrinos  (density $> 10^{12}$g/cm$^3$)
the reversed reactions (neutrino absorptions) 
may take place as well.  
The balance between neutrons and protons are determined 
through quasi-equilibrium and 
the speed of deleptonization by neutrino emissions.  

\subsection{Deuteron Formation from Nucleon-Nucleon Scattering} 
\label{sec:dformation}

The deuteron formation processes from two nucleons, (\ref{eqn:nnd}), 
(\ref{eqn:ppd}) and (\ref{eqn:pnd}), 
take place regardless of the abundance of deuterons.
The CC processes, (\ref{eqn:nnd}) and (\ref{eqn:ppd}), 
 occur in addition to 
the modified Urca processes, (\ref{eqn:nnnp}) and (\ref{eqn:ppnp}), and   
the direct Urca processes (\ref{eqn:elep}) and (\ref{eqn:posn}).
They are part of the reactions which 
determine the matter composition under quasi-chemical equilibrium.
As is well known, in cold neutron stars when
the proton fraction is small enough, 
the direct Urca process (\ref{eqn:elep}) is hindered, 
and the modified Urca process (\ref{eqn:nnnp}) is essential 
for cooling.
The processes with positrons in the initial state 
are hindered in cold neutron stars, where electrons are degenerate
and positrons are scarce.
In a non-degenerate situation where the temperature is high enough,
both the processes, (\ref{eqn:nnd}) and (\ref{eqn:ppd}),
can proceed in the supernova core  environment.

The process (\ref{eqn:pnd}) 
is a new NC process to be considered in addition to 
conventional nucleon-nucleon bremsstrahlung, (\ref{eqn:nnbrems}).  
Nucleon-nucleon bremsstrahlung with $\nu\bar{\nu}$ pair emission
is one of the main cooling mechanisms
of cold neutron stars. 
Suzuki~\citep{suz93} pointed out the importance of this process 
as a dominant source of $\nu_\mu/\bar{\nu}_\mu$ 
and $\nu_\tau/\bar{\nu}_\tau$ 
pair-creation in proto-neutron star cooling;
it is to be noted that 
$\nu_\mu/\bar{\nu}_\mu$ and 
$\nu_\tau/\bar{\nu}_\tau$ carry away energy with 
almost no energy deposition
in the heating region\citep{bet90,kot06,jan07}.
We show in the next sections that 
the additional  $\nu\bar{\nu}$ producing channel, $NN\to d  \nu\bar{\nu}$, 
can be important for the cooling of a  proto-neutron star. 
\section{Calculation of cross sections and neutrino emissivities}
\label{sec:formulation}

The Hamiltonian for low-energy semi-leptonic 
weak processes is, to good accuracy, 
given by the product of the hadron current ($J^\lambda$)
and  the lepton current ($L^\lambda$) as
\begin{eqnarray}
H_W^{CC} & = & \frac{{G_F'} V_{ud}}{\sqrt{2}}\int d\vec{x} [J^{CC}_\lambda(\vec{x})
 L_{CC}^\lambda(\vec{x}) + \mbox{h. c.}]\ , \\
H_W^{NC} & = & \frac{{G_F'} }{\sqrt{2}}\int d\vec{x} [J^{NC}_\lambda(\vec{x}) L_{NC}^\lambda(\vec{x}) + 
\mbox{h. c.} ] \ ,
\end{eqnarray}
for the CC and NC processes, respectively.
The weak coupling constant 
$G_F'=1.1803 \times 10^{-5}$ GeV$^{-2}$ is taken from~\citet{net02}, and the  
 CKM matrix element  $V_{ud} = 0.9740$ is given in~\citet{pdg12}.
The hadronic weak currents are combinations
of the vector current $V^\lambda$ and  the axial-vector current $A^\lambda$:
\begin{eqnarray}
J_\lambda^{CC} & = & V_\lambda^{\pm} - A_\lambda^{\pm} \ , \\
J_\lambda^{NC} & = & (1 - 2\sin^2\theta_W)V_\lambda^{3} - A_\lambda^{3}
 - 2\sin^2\theta_W V_\lambda^s \ . 
\end{eqnarray}
In the charged current $J_\lambda^{CC}$,
the superscript $+(-)$ denotes the isospin raising (lowering) operator.
In the neutral current $J_\lambda^{NC}$,
the superscript `3' denotes the third component of the isovector current,
while $V_\lambda^s$ is the iso-scalar vector current,
and $\theta_W$ is the Weinberg angle.
The corresponding lepton currents       
are given by
\begin{eqnarray}
L_\lambda^{CC} & = & 
\bar{\psi}_l\gamma_\lambda(1-\gamma_5)\psi_\nu
\;\;\;\;\;{\rm for\,\, CC\,\,reactions} \nonumber\\
L_\lambda^{NC} & = & 
\bar{\psi}_\nu\gamma_\lambda(1-\gamma_5)\psi_\nu
\;\;\;\;\;{\rm for\,\, NC\,\,reactions}
\end{eqnarray}
The nuclear weak currents consist of one-nucleon [impulse-approximation (IA)] 
terms and two-nucleon meson-exchange current (MEC) terms. 
In this work we consider the pion and rho meson-exchange currents.
The validity of this approach has been well tested for 
the vector current  
by comparing the model predictions with, e.g., the 
measured $n + p \rightarrow d + \gamma$ reaction data. 
As for the strength of the axial-vector exchange current, 
we follow the 
standard practice to adjust its strength 
to reproduce the experimental triton beta decay rate~\citep{sch98}. 
Detailed descriptions of the model for the nuclear currents used
in the present work are given in \citet{nsk01,net02}, 
where the basic formulation and relevant input parameters are explained.

In this article we evaluate the cross sections and emissivities 
for the DBF processes, (\ref{eqn:eled}) - (\ref{eqn:pnd}).
The cross section is given in the standard way as
\begin{eqnarray}
\sigma_{i \rightarrow f}^\alpha &\!\!\!\! =\!\!\!\! & \frac{(2\pi)^4}{s_f v_{rel}}\int 
       \prod_l \frac{d\vec{p}_{f,\, l}}{(2\pi)^3}
\delta^{(4)}(\sum_{l'} p_{f,l'} - \sum_{k'} p_{i,k'}) 
 \prod_{k''}\frac{1}{2s_{i,k''} + 1} \sum_{i,f}|<f|H_W^\alpha|i>|^2  ,
\end{eqnarray}
where $v_{rel}$ and $s_f$ are the relative velocity of the incoming particles
and the symmetry factor for the identical two nucleons
in the final(initial) state.
The summation $\sum_{i,f}$ over the spin states of the final
particles and the average over the initial spin states with spin $s_{i,\, k''}$.
The momenta of the initial (final)
particles, labeled by $k$ ($l$), are $\vec{p}_{i,k}$ ($\vec{p}_{f,\, l}$).

The emissivities for neutrinos and anti-neutrino are denoted 
by $Q_\nu$ and $Q_{\bar{\nu}}$, respectively.
They are given by integrating the transition probability over the
momenta $\vec{p}_{i,k}$ and $\vec{p}_{f,\, l}$  
with a weighting factor representing the momentum distributions:
\begin{eqnarray}
Q_{\nu(\bar{\nu})}^\alpha & = & \frac{(2\pi)^4 }{s_i s_f}
\int   \prod_{k} [\frac{d\vec{p}_{i,k}}{(2\pi)^3}]
       \prod_{l} [\frac{d\vec{p}_{f,l}}{(2\pi)^3}]
\delta^{(4)}(\sum_{l'} p_{f,l'} - \sum_{k'} p_{i,k'}) 
\nonumber\\
&\times& \omega_{\nu(\bar{\nu})} \sum_{i,f}|<f|H_W^\alpha|i>|^2 
\ \Xi \ ,
\end{eqnarray}
where $\omega_{\nu(\bar{\nu})}$ is the energy of the emitted neutrino
(anti-neutrino)
and $s_i$ and $s_f$ are symmetry factor when
two-nucleons in the initial or final state
are identical particles.
$\Xi$ represents the occupation probability of incoming
particles and the Pauli blocking of outgoing particles:   
\begin{eqnarray}
\Xi & = & \prod_{k={\rm initial\ particle}} f_k(p_k) 
\prod_{l={\rm final\ fermion} \neq \nu(\bar{\nu})} ( 1 - f_l(p_l))
\end{eqnarray}
with
\begin{eqnarray}
f_k(p_k) & = & \frac{1}{ \exp((e_k(p_k) - \mu_k)/k_B T) \pm 1} 
\end{eqnarray}
where one should use ``$+$" (``$-$") for a fermion (boson);
$e_k$ ($\mu_k$) is the energy
(chemical potential) of the particle of the $k$-th kind, and
$k_B T$ is the temperature multiplied by the Boltzmann constant.
Note that the Pauli blocking factor for the final-state
neutrino (anti-neutrino) is not included in $\Xi$.
Explicit expressions for the cross sections and emissivities for the 
various processes under consideration are given
in the Appendices.

\section{Neutrino production cross sections}\label{sec:cros}

\begin{figure}[h]
\begin{center}
\includegraphics[width=8cm]{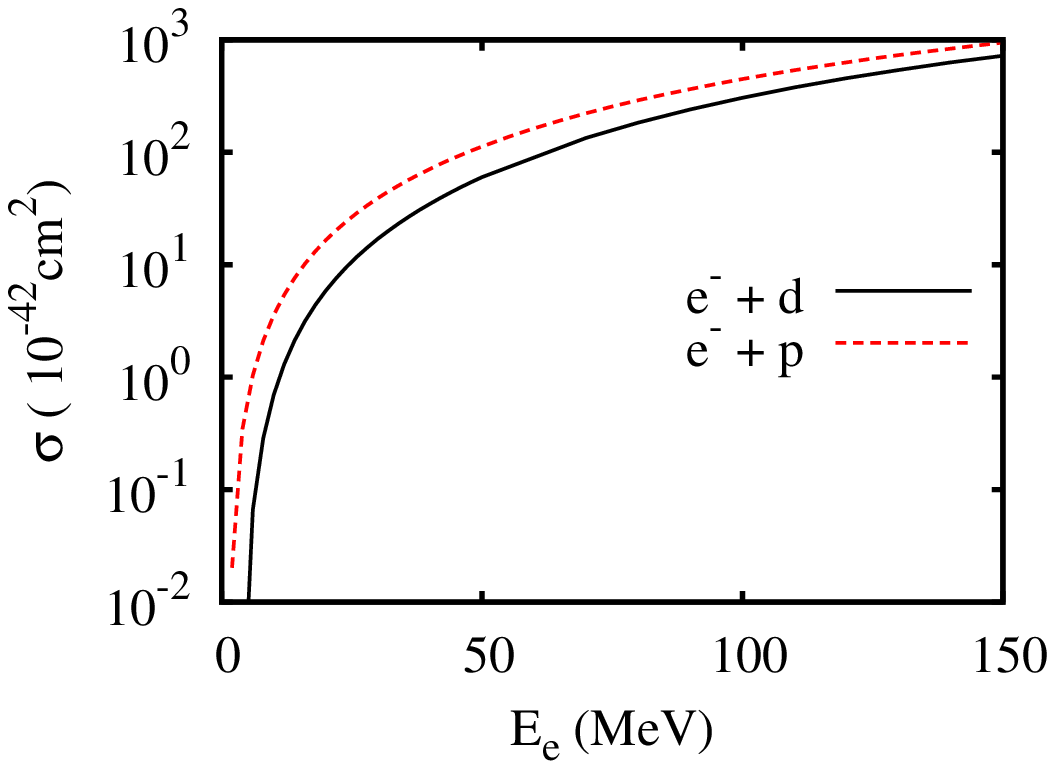}\hspace*{1cm}
\includegraphics[width=8cm]{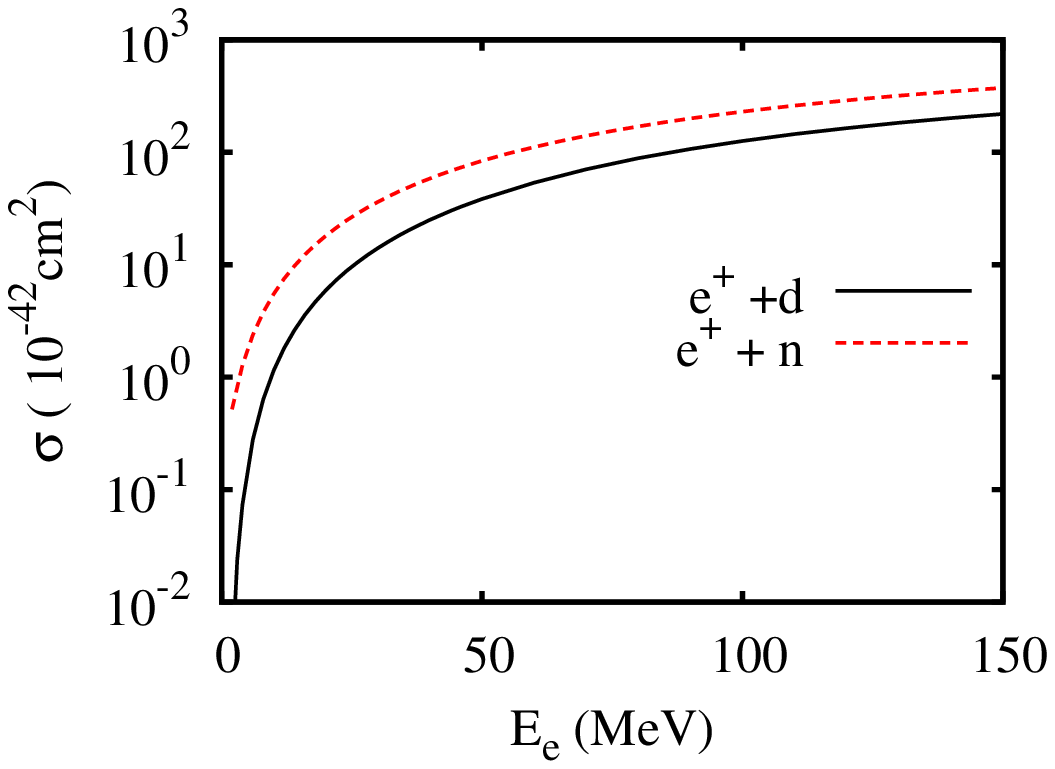}
\caption{The cross sections for $e^-/e^+$-capture on the nucleon and the deuteron
as functions of the $e^-/e^+$ energy $E_e$.
The solid and dashed curves in the left (right) panel show
the cross sections for electron (positron) capture on the deuteron
and the proton (neutron), respectively.}
\label{fig:ecap}
\end{center}
\end{figure}

In this section we discuss the cross sections 
for neutrino emissions via DBF, (\ref{eqn:eled})-(\ref{eqn:pnd}),
evaluated for initial kinetic energies up to $\sim$$100$~MeV.
Fig.~\ref{fig:ecap} shows the calculated cross sections  
for $e^-/e^+$ capture on a deuteron, 
(\ref{eqn:eled}) and (\ref{eqn:posd});
the left panel is for $e^- \!+\! d \rightarrow n \!+\! n \!+ \!\nu_e$,
and the right panel is for $e^+ \!+\! d \rightarrow p\! +\! p \!+\! \bar{\nu}_e$.
The cross sections for the two reactions are almost the same
except in the very low-energy region, 
where the $e^+$-capture cross section 
is larger than the $e^-$-capture cross sections 
due to   Q-values differences.
In the low-energy region,  $E_e < 50$~MeV, 
the dominant transition is the Gamow-Teller transition {\color{blue} to} 
the $^1S_0$ scattering state, 
and the contribution of the MEC is less than 5\%. 
It is noteworthy that, 
in the low-energy region, the cross section 
for $e^-$ ($e^+$)-capture on the deuteron
is smaller than that on the proton (neutron) by a factor larger than 3, 
mainly due to the higher threshold energy.
In the higher energy region  $E_e \sim 150$~MeV, 
the cross section for $e^-$ ($e^+$)-capture on a bound proton is almost 
comparable to that on a free proton.
The consequences of this feature in actual supernova environments 
will be discussed later in the text.

\begin{figure}[h]
\begin{center}
\includegraphics[width=10cm]{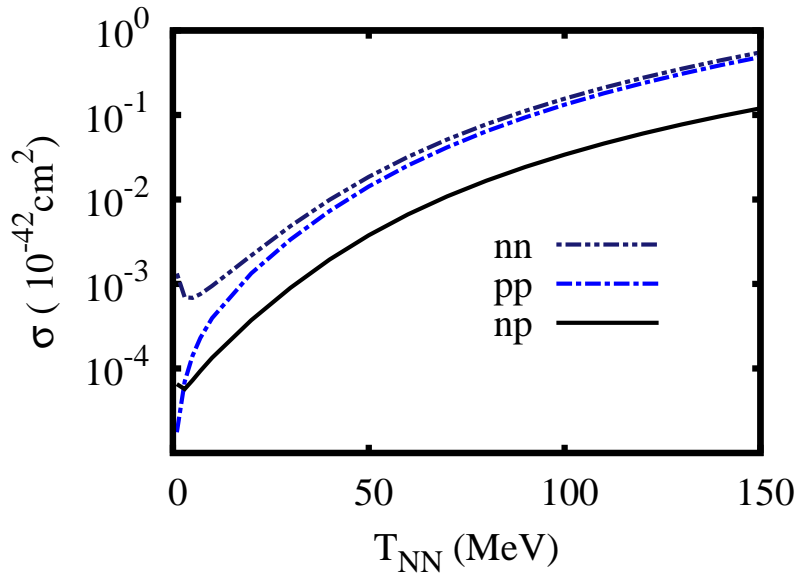}
\caption{\label{fig:nn-d} The cross sections for 
$p + n \rightarrow d + \bar{\nu}_x + \nu_x$(solid, black),
$p + p \rightarrow d + e^+ + \nu_e$(dash-dot, blue) and
$n + n \rightarrow d + e^- + \bar{\nu}_e$ (dash-two-dot, dark-blue) 
are plotted as a function of $T_{NN}$.}
\end{center}
\end{figure}

Fig.~\ref{fig:nn-d} shows 
the cross sections for neutrino production in nucleon-nucleon collision 
leading to the formation of a deuteron,
reactions (\ref{eqn:nnd})-(\ref{eqn:pnd}).
The cross sections for the CC processes are about
four times larger than the NC process,
partly due to the isospin and the symmetry factors 
for the initial identical nucleons. 
Since the processes, (\ref{eqn:nnd}) and (\ref{eqn:pnd}),
are exothermic reactions, 
their cross sections follow the $1/v$ law
in the low-energy region.
Although $pp$-fusion, (\ref{eqn:ppd}), is also an exothermic reaction,
it does not obey the $1/v$ law 
because Coulomb repulsion between the 
protons reduces the transition probability as $v$ tends to zero. 
Our result for the $pp$-fusion cross section in the keV region
agrees well with the  previous work of~\citet{sch98}.
We remark that, to calculate the neutrino processes in a supernova environment, 
it is necessary to include two-nucleon partial waves 
up to $J_{NN}<6$ ($J_{NN}$: total angular momentum)
and to consider the two-nucleon relative kinetic energy, $T_{NN}$,
up to  $T_{NN} \sim 100MeV$.
Effective field theory to describe such a kinematical region 
is just now becoming available~\citep{baru13}.
As $T_{NN}$ increases, the importance
of initial-state partial waves 
other than $^1S_0$ quickly increases, 
and furthermore,  the contribution of MEC grows and becomes 
as important as the IA contribution. 
For $T_{NN}>20$~MeV, 
the Gamow-Teller transition from the initial two nucleon $^1D_2$ state 
becomes a dominant 
transition amplitude. 
At the higher energies 
the most important MEC contribution comes from 
the $\Delta$-excitation in the axial-vector current. 
It is the tensor character of this current which produces 
a large matrix element between the 
$^1D_2$ scattering state and the deuteron S-wave.
It is notable that, even though the relevant supernova temperature   is
 $T=10-20$~MeV, 
the neutrino emissivity for $NN \rightarrow d$ at $T \sim 15$~MeV
receives the largest contribution from the energy region $T_{NN}\sim 100$~MeV.

\section{Neutrino emissivities}\label{sec:SecEmissivity}

\subsection{Supernova profiles}

In order to study the consequences of neutrino emissions due to DBF 
for the supernova-explosion mechanism,  
we calculate neutrino emissivities for a given profile of a
core-collapse supernovae,
and compare the emissivities due to DBF
with those arising from the conventional processes.  
To this end,  we consider two representative profiles of a supernova core,
Compositions I and II.

\begin{figure}[h]
\begin{center}
\includegraphics[width=5cm]{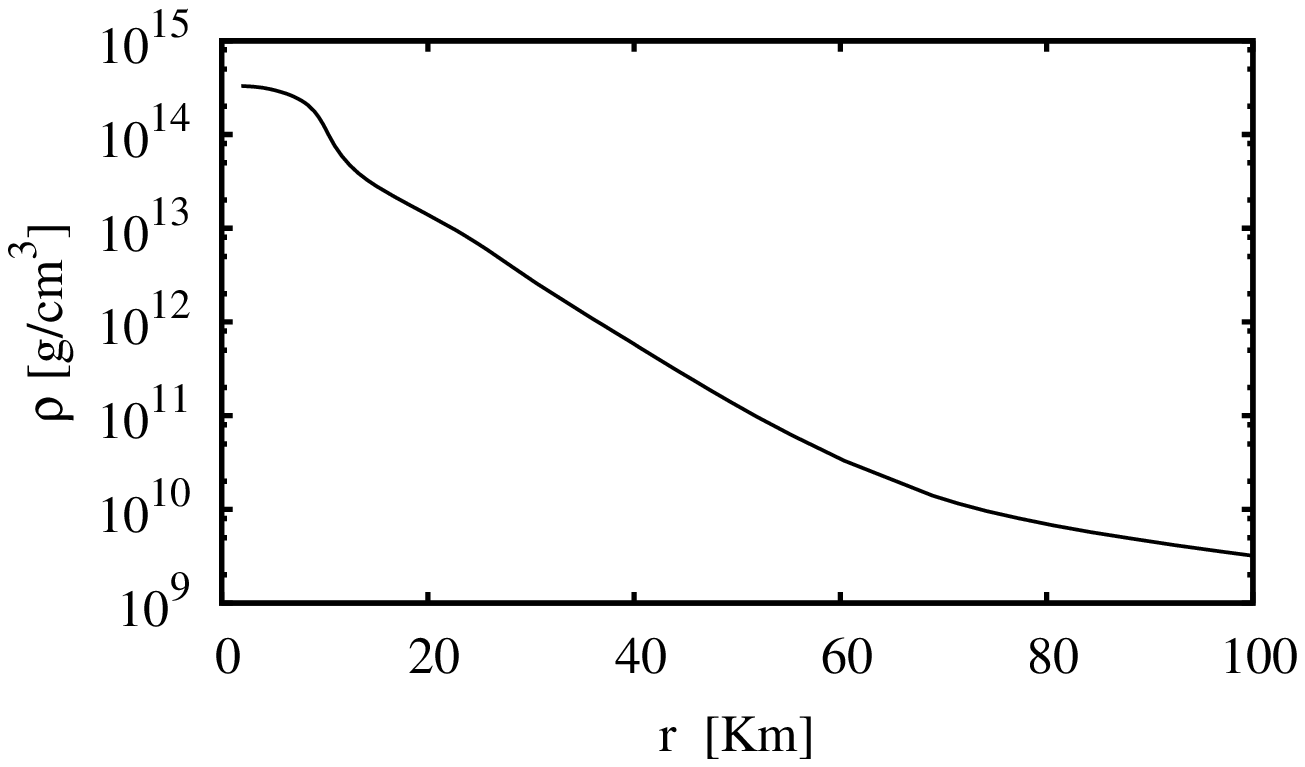}\hspace*{0.6cm}
\includegraphics[width=5cm]{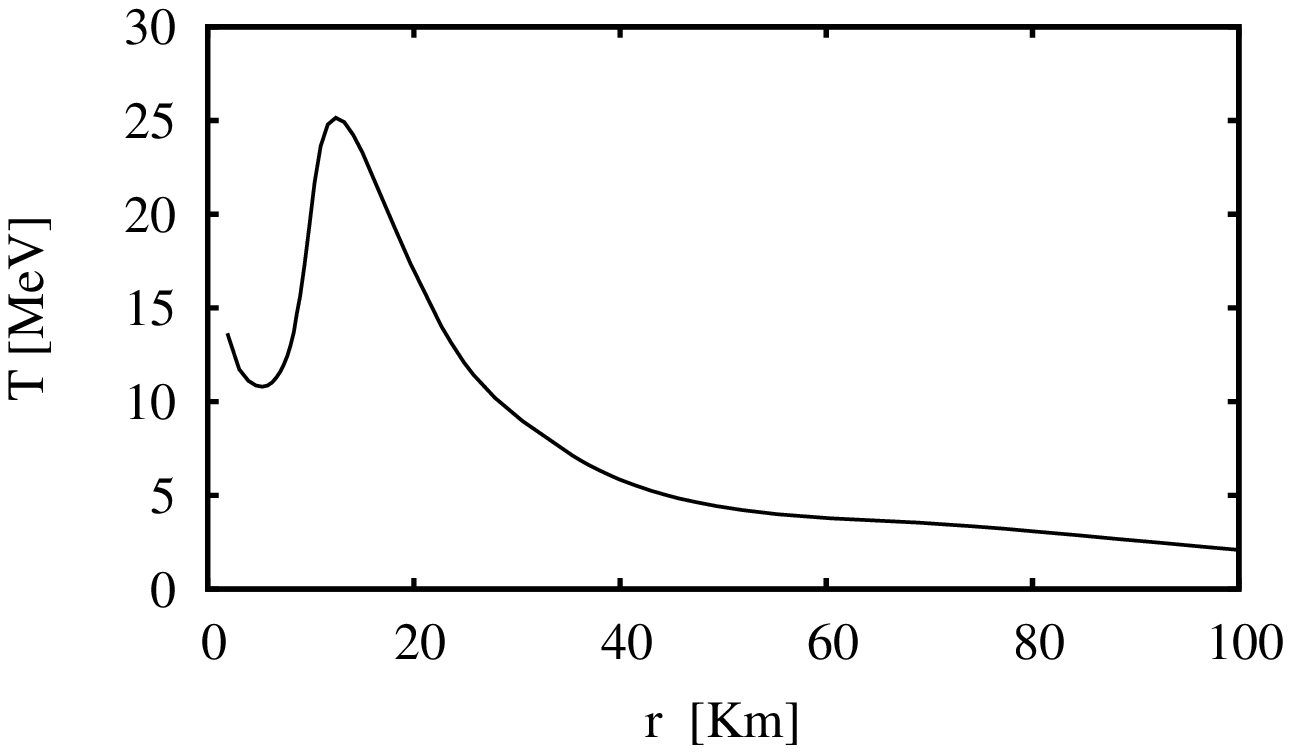}\hspace*{0.6cm}
\includegraphics[width=5cm]{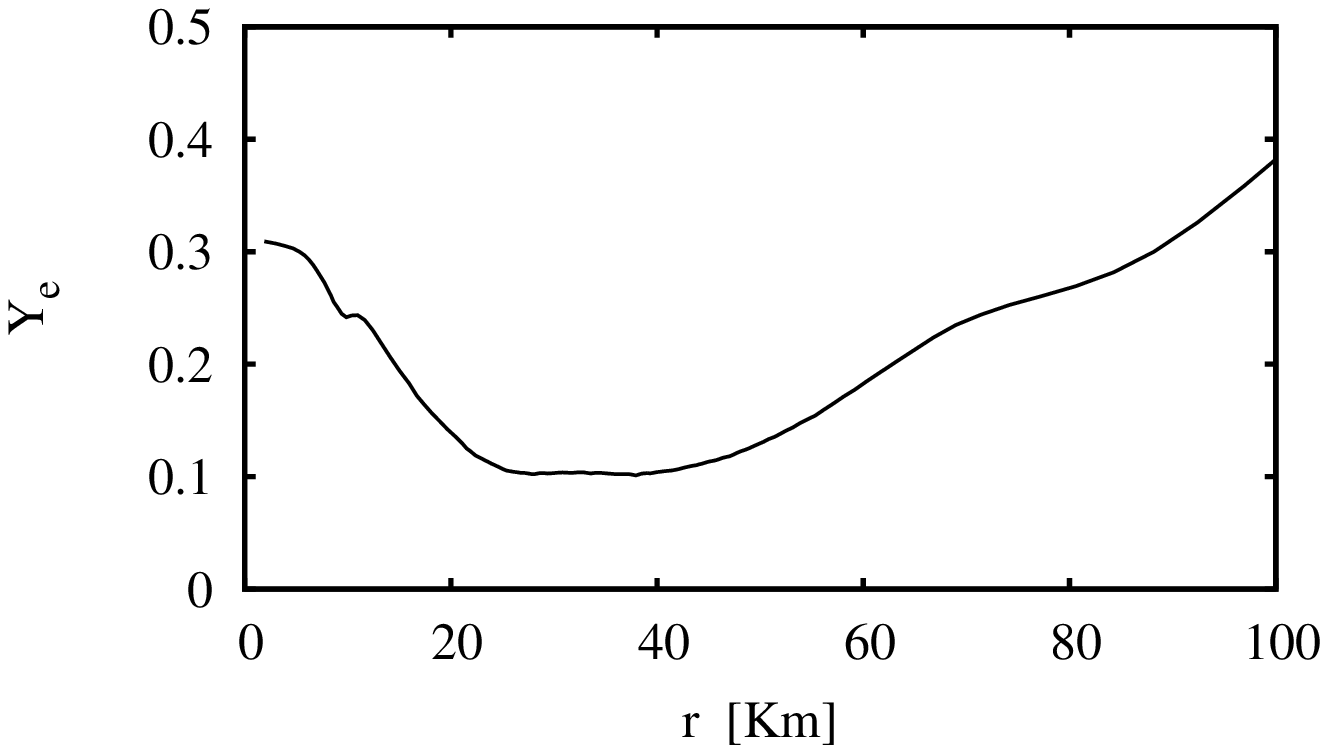}
\caption{\label{fig:profile} The density(left panel),
temperature (middle panel) and 
electron fraction (right panel) distributions
pertaining to a snapshot at 150 ms after the core bounce taken
from~\citep{sumroe08}.
The horizontal axis, $r$, is the distance from the supernova center.
}
\end{center}
\end{figure}

Composition I is the one obtained in~\citep{sum05} 
in simulating gravitational collapse and core bounce
for a 15~$M_{\odot}$ star ($M_{\odot}$: solar mass).  
This composition, which represents a typical situation of the post-bounce phase 
with a stalled shock wave,  
has been obtained from a numerical simulation 
adopting the Shen equation of state (EOS) \citep{she98a,she98b,she11}.
Composition I 
includes only nucleons, $^{4}$He and a single heavy nucleus 
in the Shen EOS.  
Fig.~\ref{fig:profile} shows 
 the  temperature ($T$) and the density ($\rho$) 
as functions of the distance $r$ from the supernova center,
pertaining to a snapshot at 150 ms after the core bounce. 


To assess the significance of the new additional emissivities due to DBF, 
we consider Composition II, 
which includes the mass fractions    
of the light elements 
obtained from the nuclear statistical equilibrium model~\citep{sumroe08}, i.e.,  
nucleons, deuterons, tritons, $^{3}$He, $^{4}$He and other nuclei are
taken into account. 
We remark that Compositions I and II share the same data 
for the profiles of  $\rho$, $T$ and  $Y_e$. 
shown in Fig.~\ref{fig:profile}.   
The nucleon chemical potentials needed to calculate the
emissivities are also taken from  the Shen EOS.  

Two regions in the profile will be discussed separately:
the surface region of a proto-neutron star
($r > 20$ km, $\rho <$ 10$^{13}$ g/cm$^{3}$)
and the inner region ($r < 20$ km, $\rho >$ 10$^{13}$ g/cm$^{3}$). 
The former corresponds to the neutrino-sphere region between the surface 
of the nascent proto-neutron star and the shock wave, 
where neutrino cooling and heating are important.  
The latter corresponds to a high density region 
 in the core of the proto-neutron star.
One can legitimately question 
the existence of free-space deuterons
in dense nuclear medium like the core region. 
Our aim here is to make a first study 
of possible influences of deuteron-like correlations
that may persist even in the core region.
Obviously our results for the core region 
obtained with the use of free-space deuterons
are of exploratory nature and should be taken as such.
The calculated emissivities are shown  
in Figs.~\ref{fig:ecap-out}-\ref{fig:nn-in-muon}. 

\subsection{Emissivity from the surface region of a proto-neutron star}

\begin{figure}[h]
\begin{center}
\includegraphics[width=7cm]{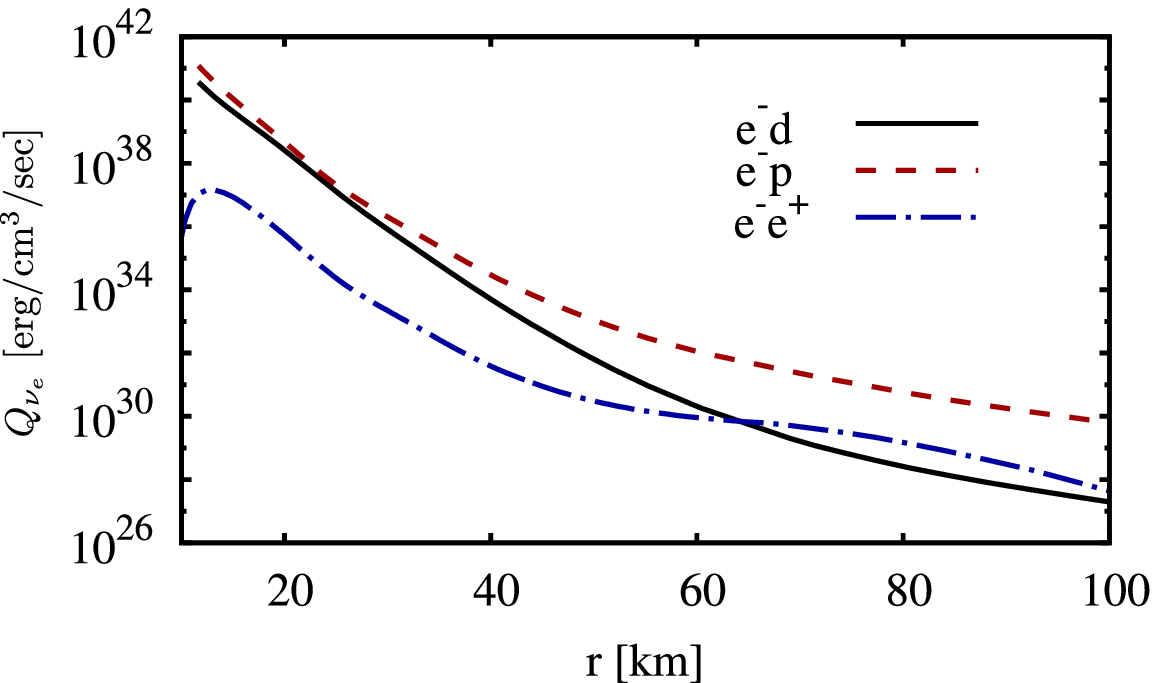}\hspace*{1cm}
\includegraphics[width=7cm]{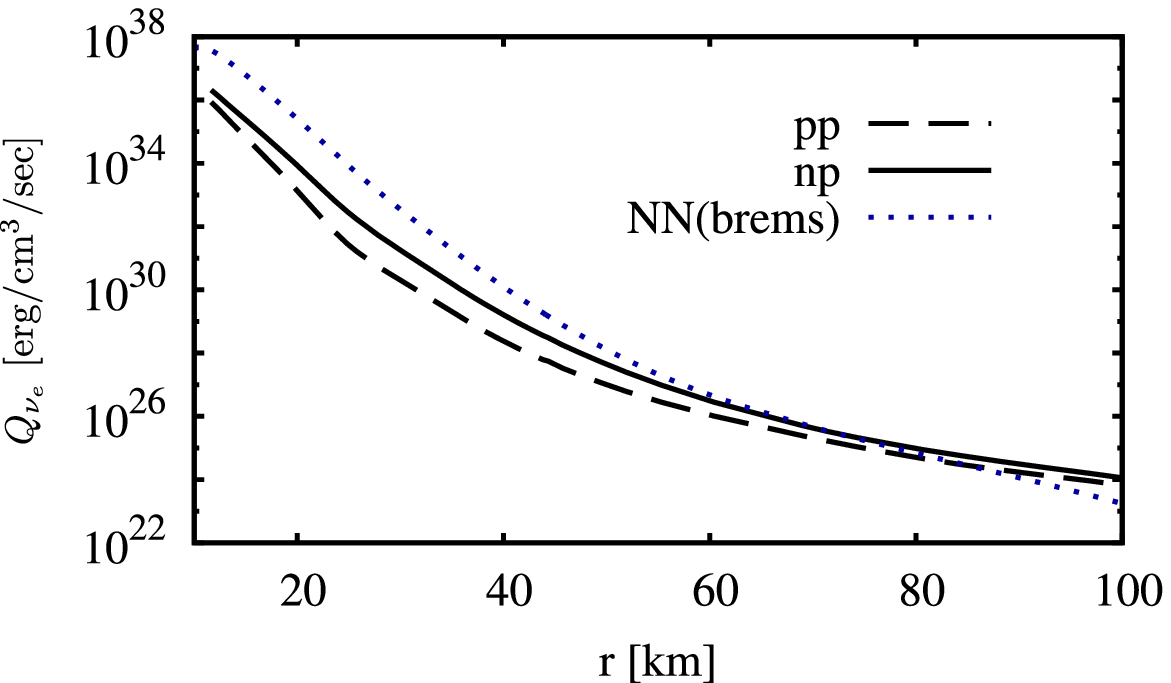}
\caption{The $\nu_e$-emissivities are shown as functions of the 
distance $r$ from the center of the supernova
evaluated with composition II except NN bremsstrahlung.
In the left panel, the neutrino emissivities 
due to $e^-$ captures on deuteron~(\ref{eqn:eled})
and proton~(\ref{eqn:elep}), and $e^+e^-$ annihilation~(\ref{eqn:epem}) 
are shown in solid, dashed and dash-dotted curves, respectively.
In the right panel,
the emisivities due to the $pp$ and $np$ fusion processes,
(\ref{eqn:ppd})  ~(\ref{eqn:pnd}),
and NN bremsstrahlung~(\ref{eqn:nnbrems})
are shown in long-dash, solid and dotted curves, respectively.
The emissivities due to the reactions (\ref{eqn:nnbrems}) and
(\ref{eqn:epem}) are taken from Ref.~\citet{sum05}. }
\label{fig:ecap-out}
\end{center}
\end{figure}

To set the stage for examining the possible influences
of $\nu_e$-emissivities due to DBF,
we first present $\nu_e$-emissivities  
due to the conventional reactions calculated for Composition II
except nucleon-nucleon bremsstrahlung (\ref{eqn:nnbrems}),
which is calculated for Composition I.
The left panel in Fig. \ref{fig:ecap-out}
shows the emissivities arising from $e^-p$ capture~(\ref{eqn:elep})
and $e^-e^+$ annihilation~(\ref{eqn:epem}),
while the right panel gives the emissivities due to
nucleon-nucleon bremsstrahlung (\ref{eqn:nnbrems}).
The figure indicates that $e^-p$ capture gives a dominant contribution,
and nucleon-nucleon bremsstrahlung (\ref{eqn:nnbrems}) and
the pair-production process (\ref{eqn:epem}) 
give only minor contributions to the emissivity.

 The left panel in Fig.~\ref{fig:ecap-out}  also
gives the neutrino emissivity due to $e^-$-capture 
on the deuteron (\ref{eqn:eled}). 
The neutrino emissivity due to $e^-$-capture 
on the deuteron is almost the same magnitude as
that on the proton~(\ref{eqn:elep}) 
around $r \approx $ 30km. In this region, the mass fraction 
of deuteron is comparable  or even larger than that of proton.
We will discuss below the 
role of including the composition of light elements
in the neutrino emissivity.


In Fig.~\ref{fig:ecap-out} (right panel) it is shown that 
the neutrino emissivities from deuteron formation
(\ref{eqn:ppd}) and (\ref{eqn:pnd}) are orders of magnitude smaller
than those from $e^-$-captures, (\ref{eqn:eled}) and (\ref{eqn:elep}),
and the pair-production process~(\ref{eqn:epem}). 
However, the neutrino emissivities from deuteron formations become 
comparable to the $\nu \bar{\nu}$ emissivity from nucleon-nucleon 
bremsstrahlung for distances closer to 100 km in the cooling region.

\begin{figure}[h]
\begin{center}
\includegraphics[width=7cm]{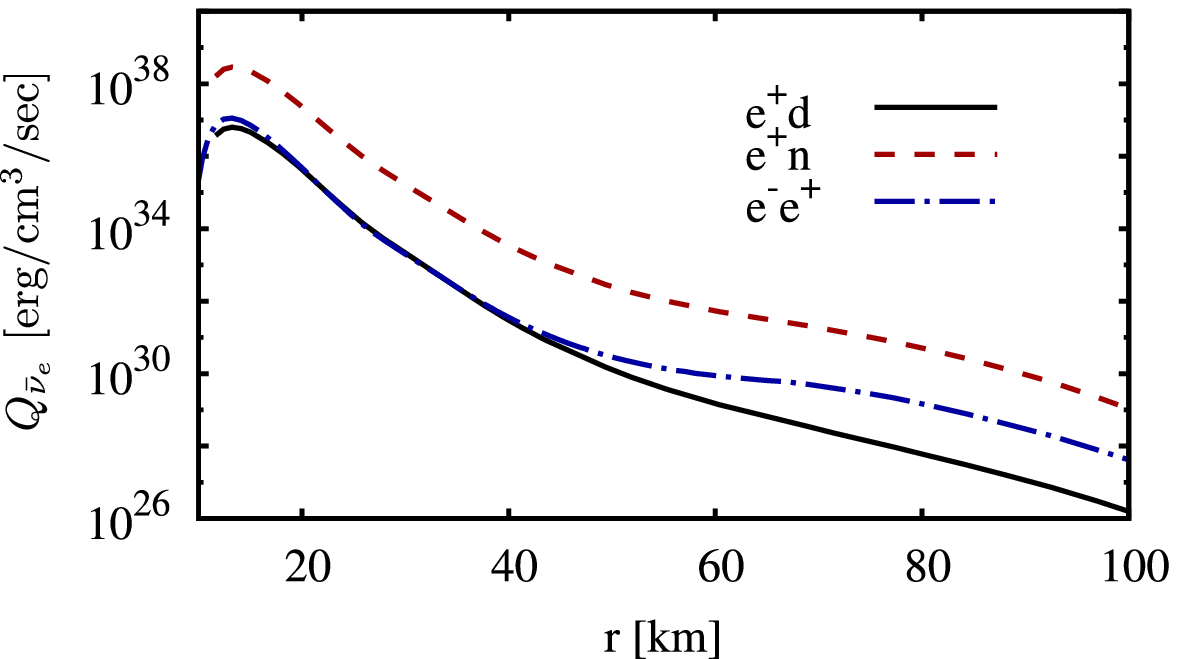}\hspace*{1cm}
\includegraphics[width=7cm]{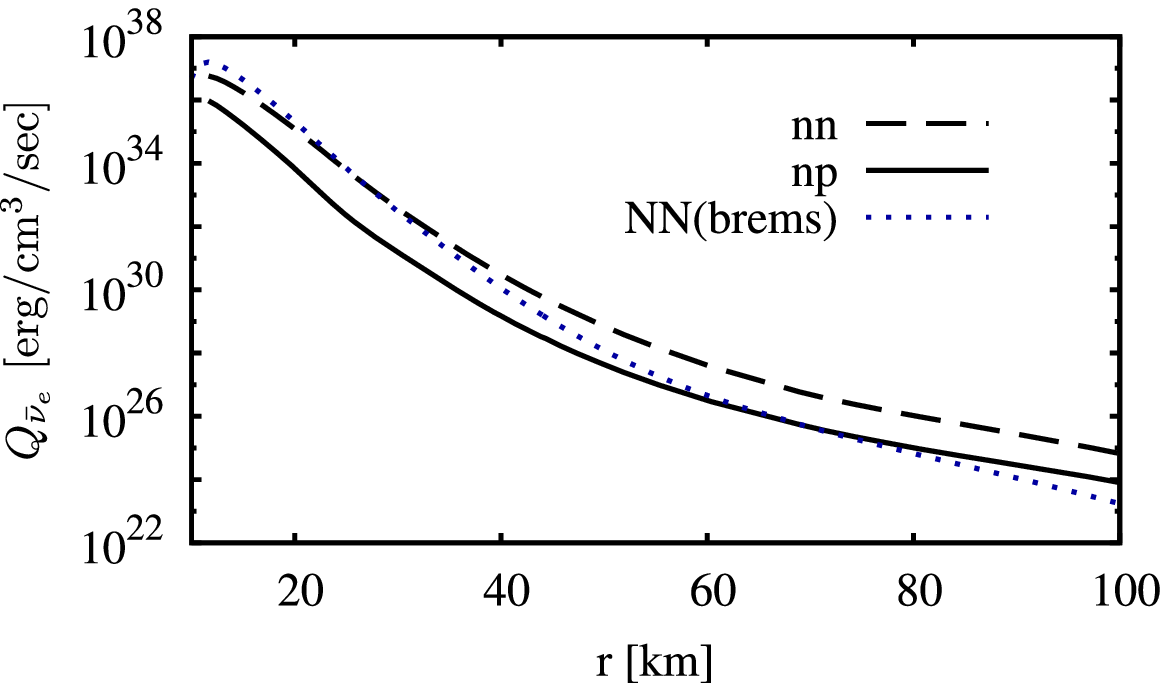}
\caption{The $\bar{\nu}_e$-emissivities evaluated with Composition II
except NN bremsstrahlung.
In the left panel, the emissivities due to $e^+$ captures on a deuteron
(\ref{eqn:posd}) and a neutron (\ref{eqn:posn}),
and $e^+e^-$ annihilation (\ref{eqn:epem}) are shown
in solid, dotted and long dashed curves, respectively.
In the right panel, the emissivities due to the $nn$ (\ref{eqn:nnd}) 
and $np$ (\ref{eqn:pnd}) fusion processes and 
NN bremsstrahlung(\ref{eqn:nnbrems}) are plotted
in dash-two-dotted, solid and two-dotted curves, respectively}
\label{fig:nn-out}
\end{center}
\end{figure}

As for the $\bar{\nu}_e$-emissivity shown in Fig.~\ref{fig:nn-out}, 
$e^+$-capture on the neutron (\ref{eqn:posn}) is dominant over the other
processes due to the very large neutron abundance as well as the relatively
large cross sections. 
As seen, the emissivity due to $e^+$-captures on the deuteron (\ref{eqn:posd}) is
smaller than those on the neutron by a factor of 10$^{2}$--10$^{3}$,
but is comparable to the pair-production process for $r < 40$ km. 
The emissions from nucleon-nucleon bremsstrahlung 
and deuteron formation are also smaller 
than $e^+$ capture on the deuteron.
We observe that the deuteron formation from
$nn$ and $np$ is comparable to that from NN bremsstrahlung 
and is significant in the outer region. 

\begin{figure}[h]
\begin{center}
\includegraphics[width=7cm]{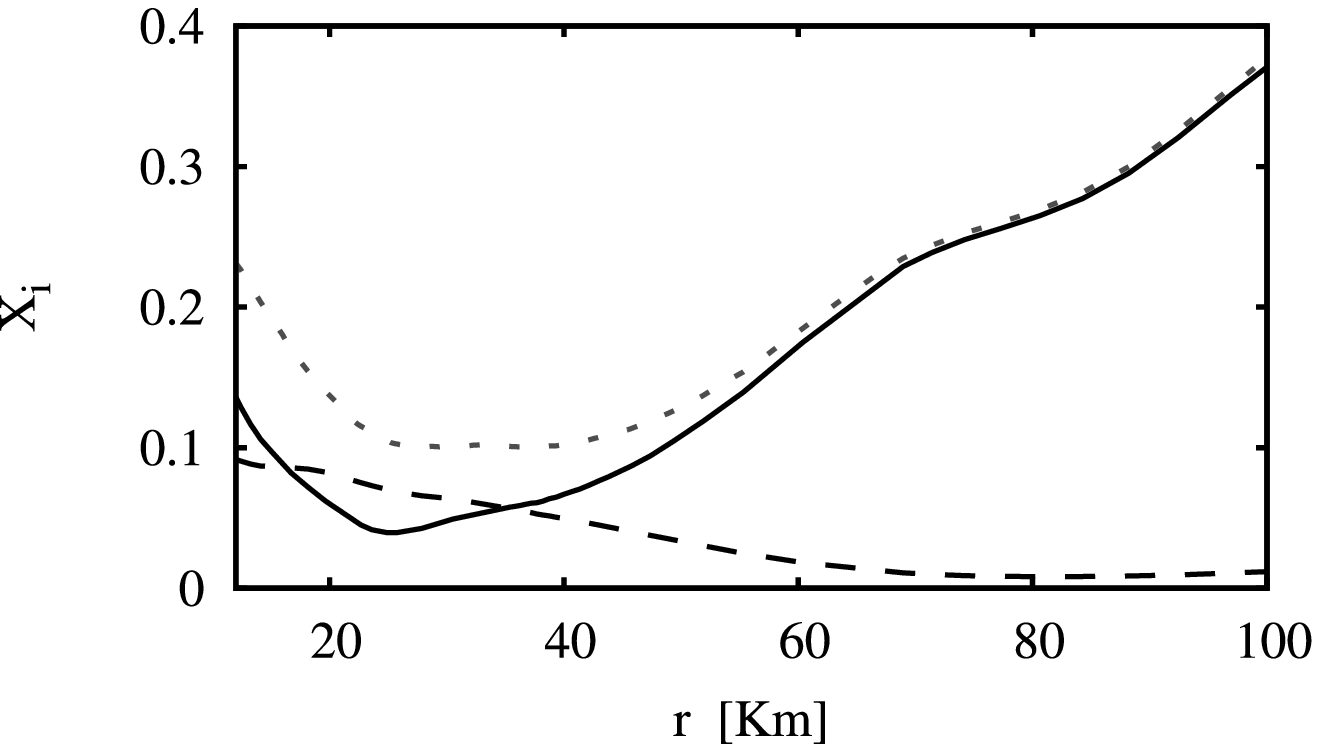} \hspace*{1cm}
\includegraphics[width=7cm]{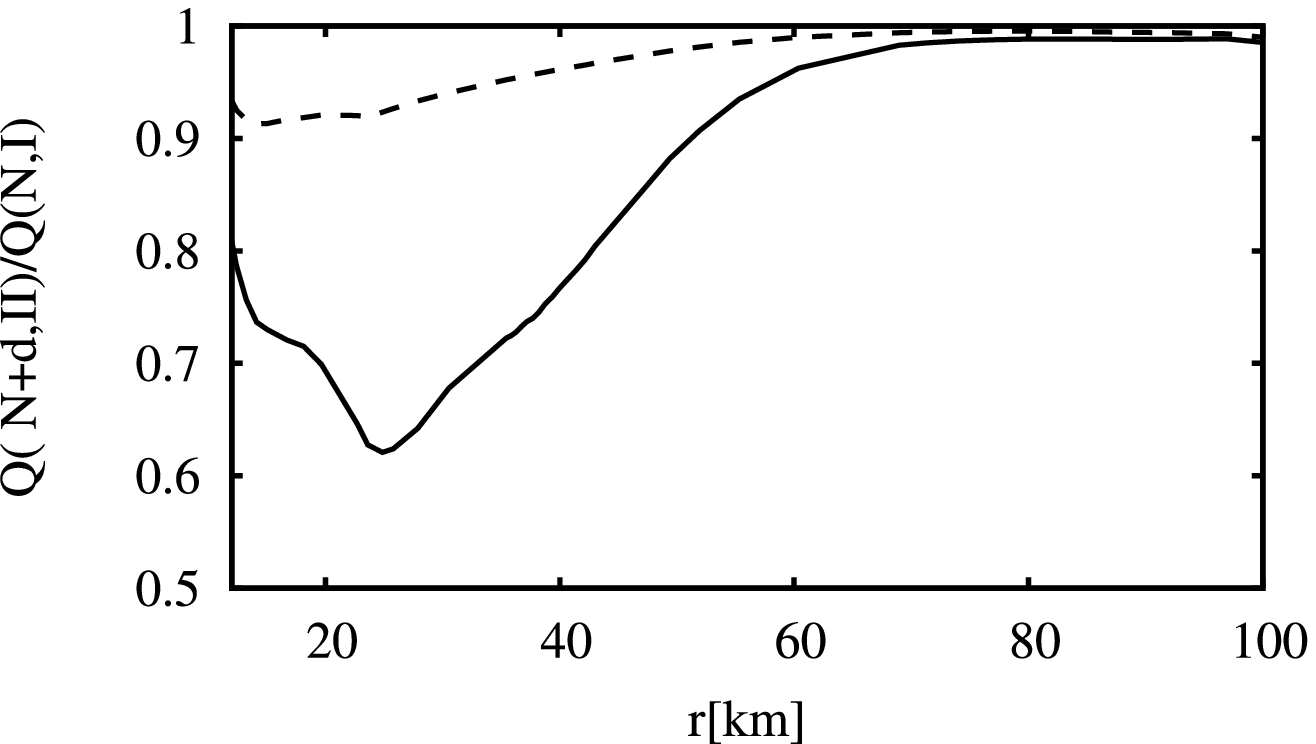}
\caption{Left panel: Mass fraction of proton (solid curve)
and deuteron (long dashed curve) for Composition II
and that of proton(short dashed curve) for Composition I.
Right panel:The ratio of the neutrino emissivity 
due to $e^-$-capture(solid curve) and
$e^+$-capture(short dashed curve) calculated for Composition II
to that calculated for Composition I.}
\label{fig:ecap-ratio}
\end{center}
\end{figure}

As we have seen in the previous section, 
the electron capture cross section of deuteron is smaller than
that of proton. This indicates that the effective neutrino emissivity {\it per proton}
via $e^-$-captures on the proton is reduced
if a substantial amount of protons in a supernova are bound in  
deuterons (and tritons).  
In fact, this is seen in Fig.~\ref{fig:ecap-ratio}.
The left panel in Fig.~\ref{fig:ecap-ratio} shows
that the free proton abundance for Composition II(solid curve) is smaller 
than that for Composition I(short dashed curve) by a factor of $\sim$2,
where deuteron(long dashed curve) is abundant.
The total emissivity due to the $e^-$-capture on protons in this region 
is effectively reduced by up to 40\% as shown
in the right panel in Fig.~\ref{fig:ecap-ratio}, where the
solid curve shows ratio of the neutrino emissivity from $e^-$-captures on the proton and
deuteron for Composition II to that from $e^-$-captures on the
proton for Composition I.
Meanwhile, as can be seen from the short dashed curve in 
the right panel of Fig.~\ref{fig:ecap-ratio},
the total anti-neutrino emissivity is only slightly affected by the deuteron
abundance, mainly due to  the  dominant abundance of the neutron in the matter.
The reduction of the total neutrino emissivity indicates that,
 in considering the neutrino emissivities 
due to electron-captures, it is important to take due account of the abundances
of deuterons and other light elements.

\begin{figure}[h]
\begin{center}
\includegraphics[width=7cm]{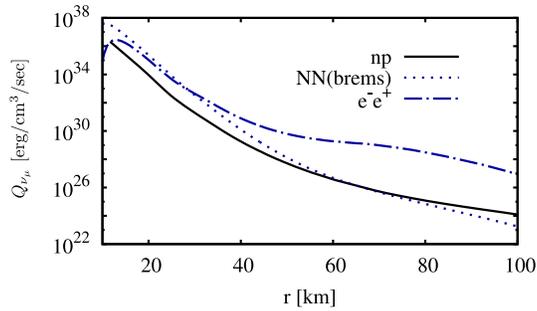}
\caption{The $\nu_\mu$-emissivities evaluated with Composition II
except NN bremsstrahlung.
The contributions of $np$ fusion, NN bremsstrahlung and $e^+e^-$ annihilation 
are shown in solid, double-dotted and dashed lines, respectively.
}
\label{fig:nn-out-muon}
\end{center}
\end{figure}

Fig.~\ref{fig:nn-out-muon} shows $\nu_\mu$-emissivities
due to $np$ fusion (deuteron formation), $NN$ bremsstrahlung and
$e^+e^-$ annihilation.   
We note that the $np$ fusion contribution is comparable
to the $NN$ bremsstrahlung contribution around $r= 60$ km,
and the former becomes more important for $r\gsim 80$ km.
In other words, 
emission of neutrino pairs through deuteron formation may
contribute to additional cooling when NN bremsstrahlung is 
an important process.

\subsection{Inner region of a proto-neutron star}

\begin{figure}[h]
\begin{center}
\includegraphics[width=7cm]{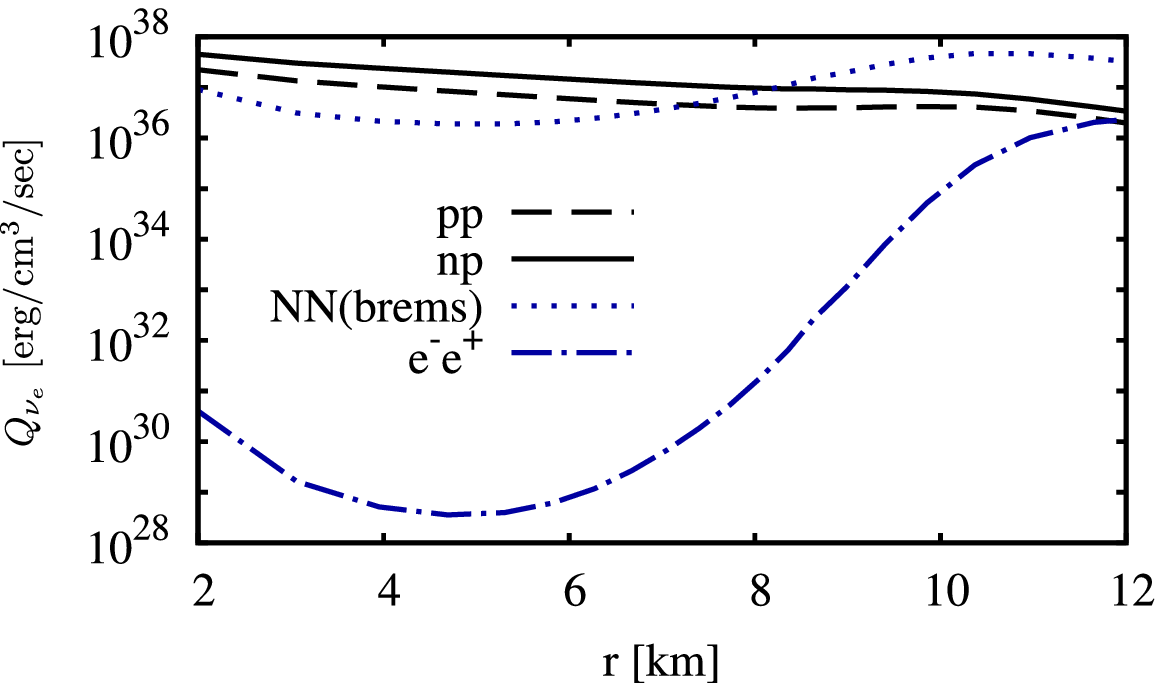}\hspace*{1cm}
\includegraphics[width=7cm]{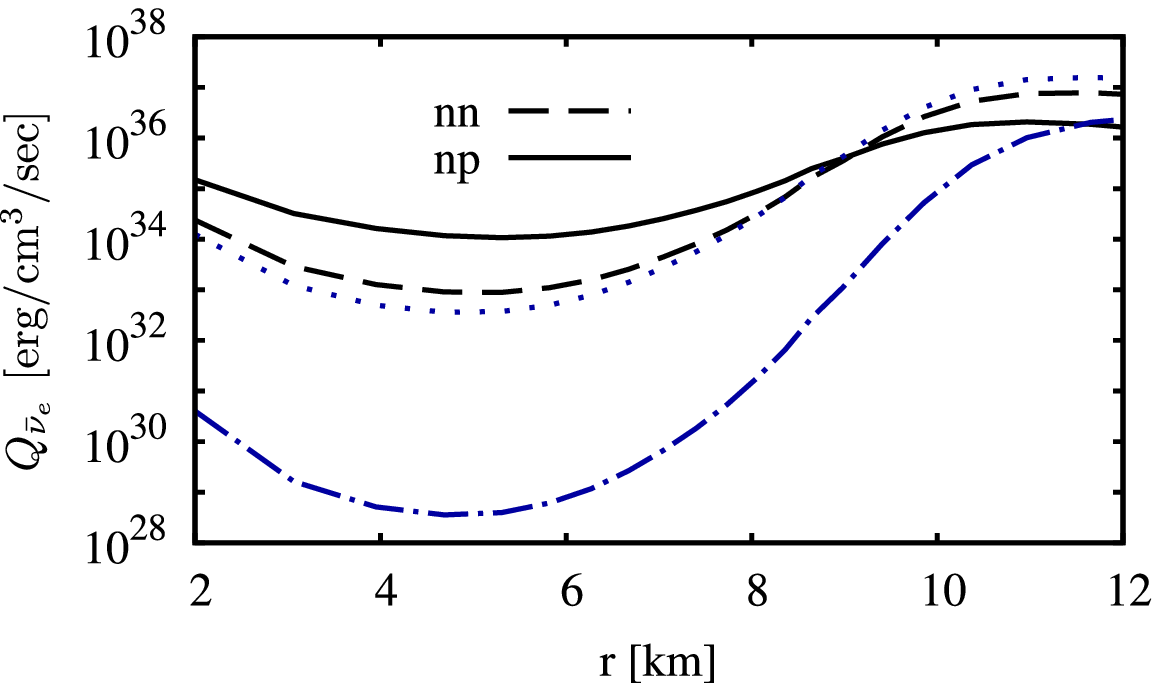}
\caption{The $\nu_e$-emissivities (left panel) and 
$\bar{\nu}_e$-emissivities (right panel)
in the inner core region  evaluated with Composition I. 
See captions for Figs. 4 and 5 for details. 
}
\label{fig:nn-in}
\end{center}
\end{figure}

\begin{figure}[h]
\begin{center}
\includegraphics[width=7cm]{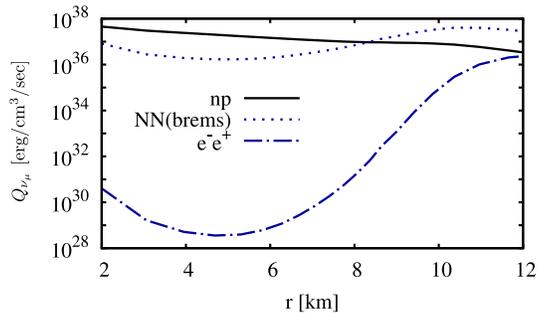}
\caption{The $\nu_\mu$-emissivities in the inner core region calculated
with  Composition I.
See the caption for Fig. 7 for details.
}
\label{fig:nn-in-muon}
\end{center}
\end{figure}

In this high density region  
the deuteron is strongly modified and is not bound. 
As mentioned earlier, 
the ``deuteron" used in our calculation  
should be regarded as a simplistic devise
to simulate possible two-nucleon 
tensor correlation in nuclear matter. 
It is hoped that the results in this section
give us some hint on whether we need to go beyond the 
mean-field nuclear matter approach and 
include possible two-nucleon correlations. 
With this caveat in mind we present the
neutrino emissivities via the ``deuteron" formation processes 
in the central part of the supernova core.
Figs.~\ref{fig:nn-in} and \ref{fig:nn-in-muon}
show the results obtained with the use of Composition I.
The graphs in Figs.~\ref{fig:nn-in} and \ref{fig:nn-in-muon}
indicate that neutrino emissions via 
``deuteron" formation 
become dominant for $\nu_e, \bar{\nu}_e$ and ${\nu}_\mu$.
Since the electrons are highly
degenerate in the core ($r < 10$ km), 
the pair process is strongly suppressed and  
nucleon-nucleon bremsstrahlung is a dominant channel
in the conventional models. 
However, the figures show that 
neutrino pair emission from the neutron-proton weak fusion
process (\ref{eqn:pnd}) 
is much larger than those from the conventional processes, in particular
for $\bar{\nu}_e$.
The reaction (\ref{eqn:pnd}) is favored by its positive Q-value 
due to the ``deuteron" binding energy, 
and by the absence of Pauli-blocking in the final state. 
Hence, neutrino emission via ``deuteron" formation 
from nucleon-nucleon scattering may play an important role
in the neutrino pair production process and  for  the  
transport of heat and leptons inside proto-neutron stars.  
It is desirable to examine further the abundance of ``deuteron" 
(n-p tensor correlations) in dense nuclear matter.

\section{Discussion and summary}\label{sec:summary}

It was pointed out in Refs.~\citet{sumroe08} and \citet{arc08}
that, in addition to deuterons, tritons can also have large abundances
in high density regions (10$^{11}$--10$^{14}$ g/cm$^{3}$),
where the electron fraction $Y_e$ is low and the temperature $T$ is high.
This suggests the possible importance of neutrino emissivities involving the triton
or ``triton" (triton-like three-nucleon correlation in dense matter).
In the present work, however,
we have not considered these effects.

Our study here is based on the spherical (1D) 
configurations of supernovae.  
It would be interesting to study neutrino emission 
with the abundance of light elements in 2D/3D profiles.  
Hydrodynamical instabilities can generate 
non-spherical distribution of matter in
the cooling region around a proto-neutron star
and the heating region behind a stalled shock wave. 
Since the density, temperature and electron fractions
can have wider ranges in 2D/3D profiles,
there may be regions of high deuteron abundance that
cannot be found in the 1D profile.
The existence of deuterons in the heating regions can contribute 
to the additional source of heating as studied by \citet{nakamura09}.  
It thus seems interesting to study the possible effects 
of the neutrino emission and absorption channels 
involving the deuteron in multi-D supernova explosion 
simulations. 
In principle, one must study effects of all neutrino processes 
by solving the neutrino transfer and hydrodynamics,
with detailed information on composition from the equation 
of state of supernova matter.
A study along this line has been recently made~\citep{furusawa13},
taking into account the neutrino processes involving neutrino absorption 
on the deuteron and the light element abundance.  

We now summarize. 
Neutrino emissions from
$e^\mp$-capture on the deuteron
and from deuteron formation in nucleon-nucleon weak-fusion processes 
have been studied as new neutrino emission mechanisms
in supernovae. 
These weak processes are evaluated with the
standard nuclear physics approach, which consists of the 
one-nucleon impulse current and two-nucleon
exchange current and nuclear wave functions 
derived from high-precision phenomenological $NN$ potentials.
It is found that the contribution of the two-nucleon 
meson-exchange current 
is only a few \% for the $e^\mp$-capture reactions, while
it can be as large as the one-nucleon current contribution
for the NN fusion reaction at higher energies. 
The consequences of these new neutrino-emission channels
have been examined for representative profiles
of core-collapse supernovae at 150 ms
after core bounce. 
The emissivity due to the
$e^\mp$ capture reaction on the deuteron is found to be smaller than 
that on the free nucleon.
Therefore, as Fig.~\ref{fig:ecap-ratio} indicates,  
the total neutrino emissivity due to electron capture on protons
and deuterons is suppressed when an appreciable amount of protons
in a supernova are bound inside deuterons. 
This results in a smaller neutrino luminosity and the lower efficiency 
of neutrino heating behind a stalled shock wave.  
Therefore, this new process may contribute unfavorably
towards a successful supernova explosions.  
It might lead to a slower speed of the deleptonization and, hence, 
a slower evolution of nascent proto-neutron stars.  
On the other hand, as seen in Figs.~\ref{fig:ecap-out} and \ref{fig:nn-out},
neutrino emission via deuteron formation 
can be comparable to nucleon-nucleon bremsstrahlung in the outer region.  
This implies that there might exist situations in which
the deuteron-formation weak processes
are the main channels for neutrino emission.  

In the inner core region, where the electrons are highly degenerate 
(high densities at low temperatures), 
pair-production via $e^-e^+$ annihilation is suppressed,  
making nucleon-nucleon bremsstrahlung a main channel 
to produce $\nu\bar{\nu}$ pairs
among the conventional processes\citep{suz93, bur06}.  
Meanwhile,  ``deuteron" formation processes in $NN$ scattering
can have large rates for $\nu_\mu$ and $\nu_\tau$ emissions,
a feature that may have significant consequences 
for the cooling of compact stars. 
Furthermore, the possible modification of the energy spectra
of $\nu_e$, $\nu_\mu$ and $\nu_\tau$ due to ``deuteron" formation
may influence supernova nucleosynthesis~\citep{woo90,yoshi04}
and the terrestrial observation of supernova neutrinos
\citep{nakazato13}.
On the other hand, the possible increase of
the $\nu_\mu$ and $\nu_\tau$ fluxes due to ``deuteron" formation
hardly affects the heating process behind a shockwave,  
because these low-energy $\nu_\mu$ and $\nu_\tau$ 
interact with stellar matter only through the NC.  
As explained earlier, the ``deuteron" here stands for
a tensor-correlated $NN$ pair that may persist even 
in dense nuclear matter.
A detailed study of deuteron-like two-nucleon tensor correlation 
in dense matter seems well warranted, but 
it is beyond the scope of our present exploratory work.  

\acknowledgments

This work was partially supported by 
JSPS KAKENHI Grant Numbers 25105010,20105004, 22540296, 24244036 and 24540273.
K. Sumiyoshi is grateful to 
G. R{\" o}pke, S. Furusawa, S. Yamada and H. Suzuki for fruitful collaborations 
on the composition of light elements in dense matter and 
the numerical simulations of supernovae. 
K.S. acknowledges the usage of the supercomputers 
at Research Center for Nuclear Physics (RCNP) in Osaka University, 
The University of Tokyo,
Yukawa Institute for Theoretical Physics (YITP) in Kyoto University,
and High Energy Accelerator Research Organization (KEK).
SXN is a Yukawa Fellow and his work is supported in part by Yukawa
Memorial Foundation.
FM is supported in part by the National Science Foundation (US) 
Grant No. PHY-1068305. 

\appendix

\section{Emissivity and cross section}

\subsection{Nuclear Matrix Elements}

The matrix element $<f|H_W|i>$ is evaluated 
by using the multipole expansion formula given in \citet{nsk01}.
The transition probability of electron(positron) capture 
$e^{\mp}(p) + i \rightarrow \nu(\bar{\nu})(p') + f$ for
the initial two nucleon state $i$($|LSJT,M>$) and the final state 
$f$($<L'S'J'T',M'|$) is written as
\begin{eqnarray}
\sum_{spin's} |<f(L'S'J'T',M');\nu(\bar{\nu})(p')|H_W^\alpha|i(LSJT,M);e^{\mp}(p)>|^2
& = & 2 (4\pi)  X^{\mp}_\alpha(f,i;p',p).\nonumber \\
\end{eqnarray}
Here $L,S,J$ and $T$ are the orbital, spin, total
angular momentum and isospin of the two-nucleon state.
We sum over all spin components of the leptons and two-nucleon states.
For $\alpha=$CC and NC reactions, $X_\alpha^{\mp}$ is given as
\begin{eqnarray}
 X_{\alpha}^{\mp}(f,i;p',p) & = &
\frac{G_F^2}{2} F_Z(E)\left( \begin{array}{c}
               V_{ud}^2 \\ 1 \end{array} \right) \sum_{J_o} [
\nonumber \\
& &   |<T_{C}^{J_o}({\cal V})>|^2  (  1+ \vec{\beta}\cdot\vec{\beta}'
    + \frac{q_0^2}{\vec{q}^2}\ 
 ( 1- \vec{\beta}\cdot\vec{\beta}'+
   2\hat{q}\cdot\vec{\beta}\hat{q}\cdot\vec{\beta}')
  - \frac{2 q_0}{q}\ \hat{q}\cdot(\vec{\beta}+\vec{\beta}')  )
\nonumber \\
& & + |<T_{C}^{J_o}({\cal A})>|^2 ( 1+ \vec{\beta}\cdot\vec{\beta}')
    + |<T_{L}^{J_o}({\cal A})>|^2 
( 1- \vec{\beta}\cdot\vec{\beta}'
   +2 \hat{q}\cdot\vec{\beta} \hat{q}\cdot\vec{\beta}') 
\nonumber \\
& & + 2 Re [<T_{C}^{J_o}({\cal A})><T_{L}^{J_o}({\cal A})>^*]\  
       \hat{q}\cdot(\vec{\beta}+\vec{\beta}')  \nonumber \\
& & + [|<T_{M}^{J_o}({\cal V})>|^2 + |<T_{E}^{J_o}({\cal V})>|^2 +
       |<T_{M}^{J_o}({\cal A})>|^2 + |<T_{E}^{J_o}({\cal A})>|^2]\ 
 \nonumber \\
& & \times 
      (1-\hat{q}\cdot\vec{\beta}\ \hat{q}\cdot\vec{\beta}')  \nonumber \\
& &  \mp 2 Re[<T_{M}^{J_o}({\cal V})><T_{E}^{J_o}({\cal A})>^* + 
              <T_{M}^{J_o}({\cal A})><T_{E}^{J_o}({\cal V})>^*]\ 
      \hat{q}\cdot(\vec{\beta}-\vec{\beta}')  ]\nonumber \\
& & .
\end{eqnarray}
Here $\vec{\beta} = \vec{p}/e(p)$ is the velocity of the lepton
with $\vec{p}$ and $\vec{p}'$ being  the momentum 
of the electron (positron) 
and neutrino, and the momentum transfer $q_\mu=p_\mu - p_\mu'$. 
$F_Z(E)$ is the Fermi function to take account of the Coulomb correction for 
the electron wave function.
The nuclear reduced matrix element $< O >=<f||O||i>$ of 
the multipole operator $O$ is defined in Eq. (60) of \citet{nsk01}, 
which includes all information of the nuclear current and nuclear wave functions.
For $e^-$-capture on the deuteron,
the above formula should be understood as
\begin{eqnarray}
<{\cal O}> & = & \sum_{L=0,2} <L'S'J'T';NN||{\cal O}||L,S=1,J=1,T=0;d>
\end{eqnarray}
where $|LSJT;d>$ is the deuteron bound state and 
$|L'S'J'T';NN>$  is the two nucleon scattering state. 

\subsection{Electron and positron capture on deuteron}

The cross section formula for the electron/positron capture reaction
$ e^{-}(p_e) + d(P_d) \rightarrow \nu_e(p_\nu) + n(p'_1) + n(p'_2)$ /
$ e^{+}(p_e) + d(P_d) \rightarrow \bar{\nu_e}(p_\nu) + p(p'_1) + p(p'_2)$ 
is given as
\begin{eqnarray}
\sigma_{e^{\mp}-cap} & = & \frac{m_N}{ 3\pi \beta}
 \int_0^{p_{\nu,max}}dp_\nu p_\nu^2 p'_{NN}
\int_{-1}^{1} d\cos\theta_{e\nu} 
\sum_{L',S',J',T'} X^{\mp}_{CC}(NN(L'S'J'T'=1),d;p_\nu, p_e). \nonumber \\
\end{eqnarray}
Here $N$ denotes neutron/proton for the electron/positron capture reaction.
We have introduced the relative momentum $\vec{p}'_{NN} = (\vec{p}'_1 - \vec{p}'_2)/2$
and the center-of-mass momentum $\vec{P}'=\vec{p}'_1 + \vec{p}'_2$ of 
the final two nucleons.  
As usual,  $P_d^2/2m_d \sim P^{'2}/4m_{N}$, where we have neglected
the difference between the center-of-mass energy 
of two nucleons and the deuteron, and we have used 
${p'}^2_{NN}/m_{N}= e_e(p_e) + m_d - p_\nu - 2m_n$.

The neutrino emissivity $Q_{\nu_e/\bar{\nu}_e}$ for electron neutrino 
or anti-neutrino is given as
\begin{eqnarray}
Q_{\nu_e/\bar{\nu}_e} & = &  \frac{m_N}{8\pi^6}
\int_0^\infty  dp_e\int_0^{p_{\nu,max}} dp_\nu  p_\nu^3 p_e^2 p'_{NN} 
 <\Xi>_{\nu_e/\bar{\nu}_e} \nonumber \\
& \times & 
 \int_{-1}^{1} d\cos\theta_{e\nu} 
\sum_{S'L'J',T'=1}X_{CC}^{\mp}(NN(L'S'J'T'=1),d;p_e,p_\nu).
\end{eqnarray}
Here $<\Xi>_{\nu_e/\bar{\nu}_e}$ is given as
\begin{eqnarray}
<\Xi>_{\nu_e/\bar{\nu}_e} & = &
 f_{e^{\mp}}(p_e) \int d\vec{P}_d
 f_d(\vec{P}_d)(1 - f_{N}(\vec{P}_d/2 + \vec{p}'_{NN}))
(1 - f_{N}(\vec{P}_d/2-\vec{p}'_{NN})).
\end{eqnarray}
The exact formula for the emissivity involves 8-dimensional integration. 
A standard approximation is introduced in order to make the 
numerical integration manageable.  
We  
factorize the angular dependence in the matrix element 
and $\Xi$ as follows 
\begin{eqnarray}
\int d\Omega_{p'_{NN}} |<f|H_W|i>|^2 \,\Xi 
\sim \int d\Omega_{p'_{NN}}|<f|H_W|i>|^2 \times
\int \frac{d\Omega_{p'_{NN}}}{4\pi} \Xi.
\end{eqnarray}


\subsection{Neutrino emission in nucleon-nucleon scattering}

The cross sections for neutrino emission in nucleon-nucleon scattering
$ N(p_1) + N(p_2) \rightarrow d(P_d) + l(p_l) + \bar{l}(p_{\bar{l}})$
 are given as
\begin{eqnarray}
\sigma_{NN-fusion} & = & \frac{2\mu_{NN}}{\pi p_{NN}}f_\alpha
 \int_0^{p_{max}}dp_l  p_l^2 p_{\bar{l}}^{2}
 \nonumber \\
& \times & \int_{-1}^{1} d\cos \theta_{l\bar{l}}
\sum_{LSJT} X_\alpha^+(d,NN(LSJT);p_{\bar{l}},p_l)
\end{eqnarray}
where $\alpha = CC$ for reactions (\ref{eqn:nnd}) and (\ref{eqn:ppd})
and $\alpha=NC$ for reaction (\ref{eqn:pnd}).
We denote the lepton momentum by $p_l$ and the anti-lepton momentum
by $p_{\bar{l}}$; 
$\vec{q}= -\vec{p}_l- \vec{p}_{\bar{l}}$ is the momentum transfer. 
We approximate the energy conservation relation
as  $e_{\bar{l}}(p_{\bar{L}}) = 
m_{N_1} + m_{N_2} + {p_{NN}}^2/(2\mu_{NN}) - e_l(p) - m_d $,
where $p_{NN}$ is  the two-nucleon relative momentum,
and $\mu_{NN}$ is the reduced mass . 
The isospin factor $f_\alpha$ is
$f_\alpha=1$ and $1/2$ for the CC and NC reactions, respectively.

The emissivity is given as
\begin{eqnarray}
Q_{\nu/\bar{\nu}} & =&  \frac{1}{4\pi^6}
 \int_0^\infty dp_{NN} \int_0^{p_{l,max}} dp_l 
p_{NN}^2 p_l^2 p_{\bar{l}} e_{\bar{l}}(p_{\bar{l}}) p_\nu
  <\Xi>_{\nu/\bar{\nu}} \nonumber \\
& \times & \int_{-1}^{1} d\cos\theta_{e\nu} 
\sum_{SLJT}X_{\alpha}^{+}(d,NN(LSJT);p_{\bar{l}},p_l)
\end{eqnarray}
The neutrino energy $p_\nu$ is either $p_l$ or $p_{\bar{l}}$ for
the neutrino or anti-neutrino emissivity, and 
\begin{eqnarray}
<\Xi>_{\nu/\bar{\nu}}  & = & F(p_l,p_{\bar{l}})
\int d\vec{P}  f_{N}(\vec{P}/2 + \vec{p}_{NN}) f_{N}(\vec{P}/2-\vec{p}_{NN}).
\end{eqnarray}
$F$ for the CC reactions,  (\ref{eqn:nnd}) and (\ref{eqn:ppd}), 
is given as
\begin{eqnarray}
 F(p_l,p_{\bar{l}}) & = & 1 - f_e(p_l) \ \ \mbox{for} \ \ (\ref{eqn:nnd})\\
                 & = & 1 - f_e(p_{\bar{l}}) \ \ \mbox{for} \ \ (\ref{eqn:ppd}),
\end{eqnarray}
while $F$ for the NC reaction, (\ref{eqn:pnd}), is given as 
\begin{eqnarray}
 F(p_l,p_{\bar{l}}) & = & 1 - f_\nu(p_l) \ \ \mbox{for} \ \ Q_{\bar{\nu}}\\
                 & = & 1 - f_{\bar{\nu}}(p_{\bar{l}}) \ \ \mbox{for} \ \ Q_\nu.
\end{eqnarray}

\bibliographystyle{apj}                       
\bibliography{apj-jour,sumi,sato}    

\end{document}